\documentstyle[preprint,aps,floats]{revtex}

\begin{document}

\setlength\baselineskip{20pt}

\preprint{\tighten\vbox{\hbox{CALT-68-2327}\hbox{hep-th/0104170}}}

\title{Renormalization Group Flows for Brane Couplings}

\author{Walter D. Goldberger\footnote{walter@theory.caltech.edu}  and Mark B. Wise\footnote{wise@theory.caltech.edu}}
\address{
\vspace{.5cm} 
California Institute of Technology, Pasadena, CA 91125}

\maketitle

{\tighten
\begin{abstract}

Field theories in the presence of branes encounter localized divergences that renormalize brane couplings.  The sources of these brane-localized divergences are understood as arising either from broken translation invariance, or from short distance singularities as the brane thickness vanishes.  While the former are generated only by quantum corrections, the latter can appear even at the classical level.  Using as an example six-dimensional scalar field theory in the background of a 3-brane, we show how to interpret such classical divergences by the usual regularization and renormalization procedure of quantum field theory.  In our example, the zero thickness divergences are logarithmic, and lead classically to non-trivial renormalization group flows for the brane couplings.  We construct the tree level renormalization group equations for these couplings as well as the one-loop corrections to these flows from bulk-to-brane renormalization effects.
\end{abstract}}
\vspace{0.7in}
\narrowtext

\newpage

\section{Introduction}

Field theory models in the presence of extended defects (``branes'') have attracted attention recently in the context of addressing the gauge hierarchy problem~\cite{ADD,RS}.  The computation of loop corrections to the effective action for such theories has been studied in~\cite{mira,georgi}, where it has been noted that quantum effects generate localized ultraviolet divergences that must be renormalized by field theory operators on the branes (other work on renormalization of field theory on singular spaces can be found in~\cite{bd}).  These ultraviolet divergences arise because in the limit of large tension, the branes break translation invariance and therefore lead to non-conservation of transverse momentum.  

Here, we consider another source of brane localized short distance divergences which come up in the renormalization of brane models.  These divergences, which arise in the limit of zero brane thickness, require renormalization even at the classical level.  They signify a breakdown of the field theory at scales at which the finite thickness of the brane cannot be neglected, and are analogous to the singularities of classical field theory, such as the ones that are found in classical electrodynamics with point sources.  While these singularities appear on brane backgrounds of codimension greater than one, they lead to particularly interesting classical effects for codimension two, since in this case the divergences are logarithmic, and therefore give rise to non-trivial renormalization group (RG) flows.  

To illustrate these effects, we consider a specific toy model in six dimensions in the vicinity of a 3-brane.  Within the context of this model, we show how to systematically account for the zero thickness classical divergences by using the usual regularization and renormalization procedure of quantum field theory (the necessity for renormalization of classical field theories with singular sources has been pointed out in~\cite{damour}).  We also construct the tree level RG equations for the brane localized couplings, as well as the one-loop corrections induced by the same type of bulk-to-brane renormalization effects considered in~\cite{mira,georgi,bd}.  

\section{The Model}

We consider Euclidean scalar field theory in a six dimensional flat space with a 3-brane.  The metric is taken to have a conical singularity:
\begin{equation}
ds^2 = \delta_{\mu\nu} dx^\mu dx^\nu + dr^2 + r^2 d\theta^2,
\end{equation}
where the brane is located at $r=0$, $0\leq\theta < 2\pi\alpha,$ with $\alpha \leq 1$, and $x^\mu$, with $\mu=0,\ldots 3$ are flat space coordinates parallel to the brane.  If gravity is included then $\alpha$ is related to the brane tension~\cite{sundrum}. Our scalar field theory is given by
\begin{equation}
\label{eq:action}
S=\int d^6 x \sqrt{g}\left[{1\over 2} (\partial\phi)^2 + {1\over 2} m^2 \phi^2 + {g_4\over 4!} \phi^4+\cdots\right]+ \int d^4 x \sum_{n=0}^\infty {\lambda_{2n}\over (2n)!} \phi^{2n}
\end{equation}
where $\cdots$ denotes a series of $\phi^{2n}$ bulk couplings and the second term includes brane localized interactions, such as a brane tension $\lambda_0$ and a brane mass $\lambda_2.$  As discussed in~\cite{georgi}, such terms must be included as counterterms for bulk-to-brane ultraviolet divergences that arise in the computation of loops with insertion of bulk interactions.  

If the brane is dynamical, the scalar $\phi$ will also couple to a set of Goldstone fields localized at $r=0$ that arise due to the breaking of translation invariance by the presence of the brane.  For simplicity, we will consider only the limit of large brane tension.  In this limit, the brane is rigid, so the backreaction on the fluctuations of the brane can be neglected, and the couplings of our scalar to the Goldstone modes are suppressed.  Note that for a cone deficit angle of $\pi,$ ($\alpha=1/2$), the conical singularity can be interpreted as a $Z_2$ orbifold fixed point.  On the orbifold, the fluctuations of the brane are projected out due to the $Z_2$ symmetry.

We will treat the bulk mass as well as the brane localized coupling $\lambda_2$ as small perturbations.  Then the scalar propagator is given by the solution of 
\begin{equation}
\Box_x D(x,x')=-\delta^4 (x^\mu-x'^\mu){\delta(r-r')\delta(\theta-\theta')\over r}.
\end{equation}
Using standard techniques, one finds
\begin{equation}
\label{eq:prop}
D(x,x')=\sum_{n=-\infty}^{\infty}\int {d^4 k\over (2\pi)^4}\int_0^\infty {q dq\over 2\pi\alpha} {e^{ik_\mu (x^\mu-x'^\mu)} e^{in(\theta-\theta')/\alpha}\over k_4^2 + q^2} J_{\left|n/ \alpha\right|}(q r) J_{\left|n/ \alpha \right|}(q r'),
\end{equation}
where $k_4^2$ is shorthand for $\delta_{\mu\nu} k^\mu k^\nu,$ and $J_\nu$ is a Bessel function of the first kind.  It is easy to check that for vanishing deficit angle ($\alpha=1$), this formula recovers the usual scalar propagator in six dimensions (see Appendix A).

\section{Classical RG equations}

Consider now the renormalization of the brane localized couplings appearing in Eq.~(\ref{eq:action}).  Besides the loop bulk-to-brane ultraviolet divergences of~\cite{georgi,bd}, there are also divergences at the classical level.  This can be seen by computing the tree level Green's functions for this theory.  For example, let us compute the corrections to the scalar propagator from inclusion of the brane mass term.  Summing the diagrams of Fig.~(1) yields
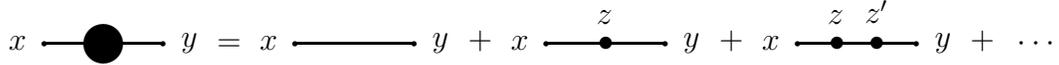
\begin{figure}
\begin{picture}(100,30)(-30,-20)
\thicklines
\put (5,0){\makebox(0,0){$x$}}
\put (15,0){\circle*{2}}
\put (15,0){\line(1,0){45}}
\put (37.5,0){\circle*{15}}
\put (60,0){\circle*{2}}
\put (70,0){\makebox(0,0){$y$}}
\put (85,-0.5){\makebox(0,0){$=$}}
\put (100,0){\makebox(0,0){$x$}}
\put (110,0){\circle*{2}}
\put (110,0){\line(1,0){45}}
\put (155,0){\circle*{2}}
\put (165,0){\makebox(0,0){$y$}}
\put (180,1){\makebox(0,0){$+$}}
\put (195,0){\makebox(0,0){$x$}}
\put (205,0){\circle*{2}}
\put (205,0){\line(1,0){45}}
\put (227.5,0){\makebox(0,0){$\bullet$}}
\put (227,10){\makebox(0,0){$z$}}
\put (250,0){\circle*{2}}
\put (260,0){\makebox(0,0){$y$}}
\put (275,1){\makebox(0,0){$+$}}
\put (290,0){\makebox(0,0){$x$}}
\put (300,0){\circle*{2}}
\put (300,0){\line(1,0){45}}
\put (315,0){\makebox(0,0){$\bullet$}}
\put (314.5,10){\makebox(0,0){$z$}}
\put (330,0){\makebox(0,0){$\bullet$}}
\put (330,12){\makebox(0,0){$z'$}}
\put (345,0){\circle*{2}}
\put (355,0){\makebox(0,0){$y$}}
\put (370,1){\makebox(0,0){$+$}}
\put (390,0){\makebox(0,0){$\cdots$}}
\end{picture}
\label{fig:prop}
\caption{Brane mass corrections to the scalar propagator.   A $\bullet$ denotes an insertion of the coupling $\lambda_2$.}
\end{figure}

\begin{eqnarray}
\label{eq:propsum}
G^{(2)}(x,y) &=& D(x,y) -\lambda_2 \int d^6 z D(x,z)\delta^2({\vec z}) D(z,y)\\
\nonumber
& &  {} + \lambda_2^2 \int d^6 z d^6 z' D(x,z)\delta^2({\vec z}) D(z,z')\delta^2({\vec z'}) D(z',y)+ \cdots,
\end{eqnarray}
where we denote coordinates transverse to the brane by a two-dimensional vector.  We will find it convenient to work in four-dimensional momentum space.  Introducing the Fourier transform of Eq.~({\ref{eq:prop}) along four-dimensional momentum,
\begin{equation}
\label{eq:momprop}
D_k ({\vec x},{\vec x'}) = \sum_{n=-\infty}^\infty\int_0^\infty {q dq\over 2\pi\alpha} {e^{in(\theta-\theta')/\alpha}\over q^2+ k_4^2} J_{\left|n/ \alpha \right|}(q r) J_{\left|n/ \alpha \right|}(q r'),
\end{equation}
Eq.~(\ref{eq:propsum}) becomes
\begin{eqnarray}
\label{eq:mom2pt}
G^{(2)}_k({\vec x},{\vec y}) &=& D_k ({\vec x},{\vec y}) - \lambda_2 D_k ({\vec x},0) D_k (0,{\vec y}) + \lambda_2^2 D_k ({\vec x},0) D_k(0,0) D_k (0,{\vec y}) + \cdots\\ 
\nonumber
&=& D_k ({\vec x},{\vec y}) - {\lambda_2\over1+\lambda_2 D_k(0,0)} D_k ({\vec x},0) D_k (0,{\vec y}).
\end{eqnarray}

In this representation for the Green's functions, momentum parallel to the brane is conserved at each vertex for a brane localized interaction.  However, due to the delta function at $r=0,$ momentum transverse to the brane is not conserved and must be integrated over each internal line for the graphs in Fig.~(1).  In Eq.~(\ref{eq:mom2pt}), this integration over two-dimensional momentum appears first at ${\cal O}(\lambda_2^2)$, and leads to the factor of $D_k(0,0)$, which is ultraviolet divergent.  We emphasize that this tree level divergence is not an artifact of our large tension limit, in which momentum appears not to be conserved due to the resistance of the brane to changes in its configuration.  Rather, it arises because we have also taken the limit in which our brane is represented by a delta function, i.e. it is infinitely thin.  In reality, the brane has internal structure at short distance, and the divergence we encounter simply reflects the fact that the field theory we wrote down in Eq.~(\ref{eq:action}) is not a valid description of the physics at these scales.  

These divergences are no different than the types of singularities that arise, for instance, in classical electrodynamics with point sources.  They can be dealt with in the same manner as the ultraviolet divergences that appear in quantum field theory, by introducing a regulator and absorbing the regulator dependence into renormalized couplings in such a way that physical quantities are regulator independent.  While the divergences described here appear on spaces with branes of codimension greater than one, they are particularly interesting for field theories on codimension two branes, such as the scalar model that we are considering here.  For codimension two, the divergences are logarithmic, leading to running couplings and RG flow even at the classical level.  To see this, regulate $D_k(0,0)$ with a momentum cutoff\footnote{It is straightforward to use dimensional regularization instead of a momentum cutoff. This would be necessary in a more realistic theory in which gauge fields or gravity are included.} $\Lambda,$ and interpret the coupling $\lambda_2$ appearing in the above series as a cutoff dependent bare coupling $\lambda_2(\Lambda)$.  Introducing a renormalized coupling $\lambda_2(\mu) = \lambda_2(\Lambda)/Z_2$ that depends on a subtraction point $\mu$, and using\footnote{Because $J_0(0)=1$ and $J_\nu(0)=0$ for $\nu>0$, only one term in the sum of Eq.~(\ref{eq:momprop}) contributes to $D_k(0,0)$.  Terms suppressed by inverse powers of $\Lambda$ have been neglected.}
\begin{equation}
D_k(0,0)={1\over 4\pi\alpha}\ln \left({\Lambda^2\over k_4^2}\right),
\end{equation}
we find
\begin{equation}
G^{(2)}_k({\vec x},{\vec y}) = D_k ({\vec x},{\vec y}) - {\lambda_2(\mu)\over 1-(\lambda_2(\mu)/ 4\pi\alpha)\ln (k_4^2/ \mu^2)} D_k ({\vec x},0) D_k (0,{\vec y})
\end{equation}
provided that we adjust 
\begin{equation}
Z_2={1\over 1-(\lambda_2(\mu)/ 2\pi\alpha)\ln(\Lambda/\mu)},
\end{equation}
which corresponds to a scheme in which only powers of $\ln(\Lambda/\mu)$ are subtracted.  Therefore, at ${\cal O}(\hbar^0)$ we have 
\begin{equation}
\label{eq:rg2}
\mu{d\lambda_2\over d\mu}={\lambda_2^2\over 2\pi\alpha},
\end{equation}
with solution 
\begin{equation}
\label{eq:l2}
\lambda_2(\mu)={\lambda_2(\mu_0)\over 1 - (\lambda_2(\mu_0)/ 2\pi\alpha)\ln(\mu / \mu_0)}.  
\end{equation}
In six dimensions $[\phi]=2$ and therefore $\lambda_2$ is a dimensionless coupling.  Eq.~(\ref{eq:l2}) indicates that for positive $\lambda_2$ this coupling increases in the ultraviolet, reaching a Landau singularity at $\mu=\mu_0 \exp(2\pi\alpha/\lambda_2(\mu_0)).$  A derivation of Eq.~(\ref{eq:rg2}) based on regulating the solution of the classical field equations derived from Eq.~(\ref{eq:action}) can be found in Appendix B.  A similar RG equation for a scalar mass term localized on a singular surface has been obtained in~\cite{massRG}.

The short distance divergences that arise in the computation of the tree level two-point function also appear at tree level in other Green's functions.  For example, the tree-level four-point function, which  can be evaluated to all orders in $\lambda_2$, is given by
\begin{eqnarray}
G_{k_1\cdots k_4}^{(4)}({\vec x_1}\cdots{\vec x_4}) &=& 
\begin{picture}(50,50)(-5,12)
\thicklines
\put(15,0){\makebox(0,0){${\vec x}_1,k_1$}}
\put(15,30){\makebox(0,0){${\vec x}_2,k_2$}}
\put(85,30){\makebox(0,0){${\vec x}_3,k_3$}}
\put(85,0){\makebox(0,0){${\vec x}_4,k_4$}}
\put(35,0){\line(1,1){15}}
\put(35,0){\circle*{2}}
\put(42.5,7.5){\circle*{7}}
\put(35,30){\line(1,-1){15}}
\put(42.5,22.5){\circle*{7}}
\put(35,30){\circle*{2}}
\put(65,0){\line(-1,1){15}}
\put(57.5,7.5){\circle*{7}}
\put(65,0){\circle*{2}}
\put(50,15){\line(1,1){15}}
\put(57.5,22.5){\circle*{7}}
\put(65,30){\circle*{2}}
\put(115,15){\makebox(0,0){$+$}}
\put(135,15){\makebox(0,0){$\cdots$}}
\end{picture}
\\
\nonumber
\\
\nonumber
&=& -\lambda_4(\Lambda) (2\pi)^4 \delta^4 (\sum_i k_i) \prod_{i=1}^4 D_{k_i}({\vec x_i},0)\left[1-{\lambda_2(\mu) D_{k_i}(0,0)\over 1 - (\lambda_2(\mu) / 4\pi\alpha)\ln ({k_i}_4^2 / \mu^2)}\right]\\
\nonumber
&=& -\lambda_4(\Lambda) Z_2^{-4} (2\pi)^4 \delta^4 (\sum_i k_i) \prod_{i=1}^4 {D_{k_i}({\vec x_i},0)\over 1 - (\lambda_2(\mu)/4\pi\alpha) \ln ({k_i}_4^2/ \mu^2)}. 
\end{eqnarray}
We define the renormalized coupling $\lambda_4(\mu)$ by $\lambda_4(\Lambda) = Z_4 \lambda_4(\mu),$ and adjust (in the same scheme used to renormalize the two-point function) $Z_4=Z_2^4.$  Then the four-point function is cutoff independent and 
\begin{equation}
\label{eq:beta4}
\mu {d\lambda_4\over d\mu} = {4\lambda_4\lambda_2\over 2\pi\alpha}.
\end{equation}
Similarly, the six-point function is given by 
\begin{eqnarray}
G_{k_1\cdots k_6}^{(6)}({\vec x}_1\cdots{\vec x}_6) &=&
\begin{picture}(50,50)(10,10)
\thicklines
\put(35,0){\line(1,1){15}}
\put(35,0){\circle*{2}}
\put(42.5,7.5){\circle*{7}}
\put(35,30){\line(1,-1){15}}
\put(42.5,22.5){\circle*{7}}
\put(35,30){\circle*{2}}
\put(65,0){\line(-1,1){15}}
\put(57.5,7.5){\circle*{7}}
\put(65,0){\circle*{2}}
\put(50,15){\line(1,1){15}}
\put(57.5,22.5){\circle*{7}}
\put(65,30){\circle*{2}}
\put(30,15){\line(1,0){40}}
\put(38,15){\circle*{7}}
\put(62,15){\circle*{7}}
\put(30,15){\circle*{2}}
\put(70,15){\circle*{2}}
\put(90,15){\makebox(0,0){$+$}}
\put(130,15){\line(-1,1){15}}
\put(130,15){\line(-1,-1){15}}
\put(150,15){\line(1,1){15}}
\put(150,15){\line(1,-1){15}}
\put(110,15){\line(1,0){60}}
\put(118,15){\circle*{7}}
\put(140,15){\circle*{7}}
\put(162,15){\circle*{7}}
\put(122.5,7.5){\circle*{7}}
\put(122.5,22.5){\circle*{7}}
\put(157.5,22.5){\circle*{7}}
\put(157.5,7.5){\circle*{7}}
\put(110,15){\circle*{2}}
\put(170,15){\circle*{2}}
\put(115,0){\circle*{2}}
\put(115,30){\circle*{2}}
\put(165,0){\circle*{2}}
\put(165,30){\circle*{2}}
\put(190,15){\makebox(0,0){$+$}}
\put(216,15){\makebox(0,0){perms.}}
\put(241,15){\makebox(0,0){$+$}}
\put(255,15){\makebox(0,0){$\cdots$}}
\put(30,0){\makebox(0,0){${}_1$}}
\put(25,15){\makebox(0,0){${}_2$}}
\put(30,30){\makebox(0,0){${}_3$}}
\put(71,30){\makebox(0,0){${}_4$}}
\put(76,15){\makebox(0,0){${}_5$}}
\put(71,0){\makebox(0,0){${}_6$}}
\put(110,0){\makebox(0,0){${}_1$}}
\put(105,15){\makebox(0,0){${}_2$}}
\put(110,30){\makebox(0,0){${}_3$}}
\put(171,30){\makebox(0,0){${}_4$}}
\put(176,15){\makebox(0,0){${}_5$}}
\put(171,0){\makebox(0,0){${}_6$}}
\end{picture}
\\
\nonumber 
\\
\nonumber
&=& (2\pi)^4 Z_2^{-6}\delta^4 (\sum_i k_i)\prod_{i=1}^6 {D_{k_i}({\vec x}_i,0)\over 1-(\lambda_2(\mu)/4\pi\alpha)\ln (k_{i4}^2/\mu^2)}
\\
\nonumber
& & \times \left[-\lambda_6(\Lambda) + \lambda_4(\Lambda)^2\sum_q {Z_2^{-1} D_q(0,0)\over 1-(\lambda_2(\mu)/4\pi\alpha)\ln(q_4^2 / \mu^2)}\right]
\\
\nonumber
&=&(2\pi)^4\delta^4(\sum_i k_i)\prod_{i=1}^6 {D_{k_i}({\vec x_i},0)\over 1 - (\lambda_2(\mu) / 4\pi\alpha) \ln ({k_i}_4^2 / \mu^2)} \\
\nonumber
& & {}\times \left[-\lambda_6(\mu) - {\lambda_4(\mu)^2} \sum_q {\ln(q_4^2 / \mu^2)\left/4\pi\alpha\right.\over 1-(\lambda_2(\mu)/4\pi\alpha)\ln(q_4^2 / \mu^2)}\right]
\end{eqnarray}
where $q$ is the four-momentum going through the internal line in the second graph in the figure, which is fixed in terms of the external momenta.  The sum over $q$ is over the momentum in this graph as well as the other nine permutations not shown in the figure.  In this expression, the renormalized coupling $\lambda_6 (\mu)$ is related to the bare coupling by $\lambda_6(\Lambda)=Z_6\lambda_6(\mu),$ with
\begin{equation}
Z_6  = Z_2^6 + {5 \choose 2}{\lambda_4(\mu)^2\over 2\pi\alpha \lambda_6(\mu)} Z_2^7\ln\left(\Lambda\over\mu\right),
\end{equation} 
leading to the RG equation for $\lambda_6$
\begin{equation}
\mu{d\lambda_6\over d\mu} = {6 \lambda_6 \lambda_2\over 2\pi\alpha} + {5\choose 2} {\lambda_4^2\over 2\pi\alpha}.
\end{equation}

In this and the previous examples, the beta functions, computed to all orders in $\lambda_2,$ coincide with those that would have been obtained by keeping only terms with two vertex insertions in the expansions for the Green's functions.  This is because such graphs are the only sources of tree level divergences that are single powers of $\ln\Lambda$.  Knowledge of the exact ${\cal O}(\hbar^0)$ coefficient of this log then determines the tree level beta functions to all orders~\cite{ramond}.  We can also immediately write the full tree level beta functions for the other brane couplings appearing in Eq.~(\ref{eq:action}).  Keeping only terms with divergences that are single powers of $\ln\Lambda$, the Green's functions are:
\begin{eqnarray}
G^{(4k)}&\sim&-\lambda_{4k}(\Lambda) + \sum_{j=1}^k {4k\choose 2j-1} {\lambda_{2j}\lambda_{4k-2j+2}\over 2\pi\alpha}\ln\left({\Lambda\over\mu}\right) + \cdots,\\
G^{(4k+2)}&\sim& -\lambda_{4k+2}(\Lambda) + \sum_{j=1}^k {4k+2\choose 2j-1} {\lambda_{2j}\lambda_{4k-2j+4}\over 2\pi\alpha}\ln\left({\Lambda\over\mu}\right) + {4k+2\choose 2k+1} {\lambda_{2k+2}^2\over 4\pi\alpha}\ln\left({\Lambda\over\mu}\right)+\cdots,
\end{eqnarray}
where in the first line $k$ runs over $k=1,2,\cdots$ and in the second equation $k=0,1,\cdots$.  The combinatoric factors count the number of distinct ways of assigning momentum labels to the external lines in the graphs.  Note that for $\lambda_{4k+2}$, the combinatorics is slightly different due to the possibility of having a graph with two factors of $\lambda_{2k+2}$.  Introducing the renormalized couplings $\lambda_{2n}(\Lambda)=Z_{2n}\lambda_{2n}(\mu)$, and choosing $Z_{2n}$ to cancel the logs of $\Lambda$,  we find 
\begin{eqnarray}
\mu{d\lambda_{4k}\over d\mu} &=& \sum_{j=1}^k {4k\choose 2j-1} {\lambda_{2j}\lambda_{4k-2j+2}\over 2\pi\alpha},\\
\mu{d\lambda_{4k+2}\over d\mu} &=& \sum_{j=1}^k {4k+2\choose 2j-1} {\lambda_{2j}\lambda_{4k-2j+4}\over 2\pi\alpha} + {4k+2\choose 2k+1} {\lambda_{2k+2}^2\over 4\pi\alpha}.
\end{eqnarray}

Because the equation for $\lambda_{2n}$ only involves the couplings $\lambda_{2m}$ with $m\leq n,$ it can be easily solved by iteration.  Given the solution for $\lambda_2(\mu)$ we construct the RG flow for $\lambda_4(\mu)$ by noting from Eq.~(\ref{eq:beta4}) that $\lambda_4(\mu)\lambda_2(\mu)^{-4}$ is an RG invariant.  Then the equation for $\lambda_6$ can be written as
\begin{equation}
{d\over d\lambda_2}(\lambda_6 \lambda_2^{-6})= {5\choose 2}\left({\lambda_4\over \lambda_2^4}\right)^2
\end{equation}
so that 
\begin{equation}
\lambda_6(\mu)\lambda_2(\mu)^{-6} = \lambda_6(\mu_0)\lambda_2(\mu_0)^{-6} + 10\left({\lambda_4(\mu)\over\lambda_2(\mu)^4}\right)^2 (\lambda_2(\mu)-\lambda_2(\mu_0)),
\end{equation}
and similarly for larger $n.$  

\section{One-Loop Corrections}

Besides the tree level RG flows just considered, there are other corrections to the brane beta functions due to loop effects involving insertions of both the bulk and brane couplings.  Loop diagrams with only brane couplings cannot give rise to any further logarithmic divergences than those obtained already at tree level.  To see this, note that an $L$ loop diagram contributing to the renormalization of the $\lambda_{2n}$ vertex with $N_{2m}$ insertions of $\lambda_{2m}$ vertices ($m=1\ldots\infty$) is proportional to a product of coupling constants that has mass dimension
\begin{equation}
d=\sum_m N_{2m}[\lambda_{2m}],
\end{equation}
where $[\lambda_{2m}]=4-4m.$  Using the relations
\begin{eqnarray}
L &=& I-\sum_m N_{2m} +1,\\
2I &=& \sum_m 2m N_{2m}-E, 
\end{eqnarray}
with $I$ the number of internal lines and $E=2n$ the number of external lines, we see that $d=4-4n-4L.$  To get a contribution to the beta function, this diagram needs to be logarithmically divergent.  This occurs when $d=[\lambda_{2n}]=4-4n,$ which happens precisely at tree level.  Thus to obtain loop corrections to the RG equations one must include insertions of the bulk couplings in loops.  We now turn to the calculation of some of these bulk-to-brane renormalization effects.

For simplicity, we will consider the field theory on a space with deficit angle\footnote {This space can also be thought of as a $Z_2$ orbifold.} $\pi$, {\it i.e.} $\alpha=1/2$.  Then the scalar propagator simplifies to a sum of two terms, the usual scalar propagator and an image charge contribution (see Appendix A)
\begin{equation}
\label{eq:image}
D(x,x') = \int {d^6 k \over (2\pi)^6} {1\over k^2}e^{i k_\mu (x-x')^{\mu}}\left(e^{i {\vec k}\cdot ({\vec x}-{\vec x}')} + e^{i {\vec k}\cdot({\vec x} + {\vec x}')}\right).
\end{equation}
In order to see what types of divergences arise from loops with insertions of bulk couplings, we compute one-loop quantum corrections to the effective action.  We will consider only the effects of the bulk $\phi^2$ and $\phi^4$ couplings.  Inclusion of the higher powers of $\phi$ is straightforward.  For a term in the effective action to give a logarithmically divergent contribution to the renormalization of $\lambda_{2n}$, it must be constructed from insertions of bulk couplings whose product has mass dimension $[\lambda_{2n}]$.  Then at one-loop, the relevant terms ({\em i.e.} the terms that diverge like a single power of $\ln\Lambda$ and therefore contribute to the RG equations) are
\begin{eqnarray}
\label{eq:eff}
S_{eff} &=& 
\begin{picture}(100,15)
\thicklines
\put(0,0){\makebox{$S_{cl}$}}
\put(17,0){\makebox{$-$}}
\put(50,3){\circle{20}}
\put(34,3){\makebox(0,0){$x$}}
\put(40,3){\circle*{3}}
\put(60,3){\circle*{3}}
\put(66,3){\makebox(0,0){$y$}}
\put(74,0){\makebox{$-$}}
\put(100,-7){\line(2,-1){15}}
\put(100,-7){\line(-2,-1){15}}
\put(100,3){\circle{20}}
\put(100,13){\circle*{3}}
\put(100,20){\makebox(0,0){$y$}}
\put(100,-13){\makebox(0,0){$x$}}
\put(100,-7){\circle*{3}}
\put(115,0){\makebox{$-$}}
\put(132,3){\makebox(0,0){$x$}}
\put(140,3){\line(-1,2){7.5}}
\put(140,3){\line(-1,-2){7.5}}
\put(160,3){\line(1,2){7.5}}
\put(160,3){\line(1,-2){7.5}}
\put(150,3){\circle{20}}
\put(140,3){\circle*{3}}
\put(160,3){\circle*{3}}
\put(168,3){\makebox(0,0){$y$}}
\put(173,0){\makebox{$+\cdots$}}
\end{picture}
 \\
\nonumber\\
\nonumber &=& S_{cl} - {m^4\over 4}\int d^6 x d^6 y D(x,y)^2 - {1\over 2!}{g_4 m^2\over 2}\int d^6 x d^6 y D(x,y)^2 \phi(x)^2\\
\nonumber & & {} - {1\over 4!}{3 g_4^2\over 2}\int d^6 x d^6 y \phi(x)^2 D(x,y)^2 \phi(y)^2 + \cdots,
\end{eqnarray}
where an external line going into a vertex at $x$ denotes an insertion of $\phi(x).$  In this expression, the second, third and fourth terms contribute to the brane tension $\lambda_0$, the brane mass $\lambda_2$ and the coupling $\lambda_4$ respectively.  Since we are only after the counterterms for the brane $\phi^{2n}$ couplings, we can take $\phi(x)=\mbox{constant}.$  Then all the integrals in Eq.~(\ref{eq:eff}) are identical
\begin{eqnarray}
\label{eq:int}
\int d^6 x d^6 y D(x,y)^2 &=& \int d^4 x \int {d^6 k\over (2\pi)^6} {d^6 q\over (2\pi)^6}{1\over k^2}{1\over q^2} (2\pi)^4 \delta^4(k-q)\\
\nonumber & & {}\times \int {d^2 {\vec x}} {d^2 {\vec y}}\left[e^{i {\vec k}\cdot ({\vec x}-{\vec x}')} + e^{i {\vec k}\cdot({\vec x} + {\vec x}')}\right]\left[e^{i {\vec q}\cdot ({\vec x}-{\vec x}')} + e^{i {\vec q}\cdot({\vec x} + {\vec x}')}\right]\\
\nonumber &=& {1\over 4} \cdot 2\int d^4 x\int {d^4 k\over (2\pi)^4} {d^2{\vec k}\over (2\pi)^2} {d^2 {\vec q}\over (2\pi)^2}{1\over k_4^2 + {\vec k}^2} {1\over k_4^2 + {\vec q}^2}\\
\nonumber  & & {}\times \left[2 (2\pi)^2\delta^2({\vec k}+{\vec q})\int d^2 {\vec x} + (2\pi)^2 \delta^2({\vec k}+{\vec q}) (2\pi)^2\delta^2({\vec k}-{\vec q})\right],
\end{eqnarray}
where the factor of $1/4$ in the second line reflects the fact both integrals over two-dimensional position run over only the half-plane.  Regulating the four-dimensional and two-dimensional momentum integrals with an ultraviolet cutoff\footnote{Here, we choose the same cutoff $\Lambda$ that we used in the previous section to regulate $D_k(0,0)$.  This choice reflects the assumption that both the zero thickness and the large tension divergences are resolved by physics at similar scales, of order the ultraviolet cutoff $\Lambda.$} $\Lambda,$
\begin{equation}
\label{eq:result}
\int d^6 x d^6 y D(x,y)^2 = \int d^4 x{1\over 64\pi^2}\ln\left({\Lambda\over\mu}\right) + \int d^6 x {\ln 2\over 64\pi^3}\Lambda^2,
\end{equation}
where we have introduced a subtraction scale $\mu.$  Note that the brane localized ultraviolet divergences encountered here are different in nature than the classical singularities discussed in the previous section.  In this case, the divergence proportional to four-dimensional volume arises because the brane at $r=0$ induces a spacetime geometry that breaks translation invariance and leads to non-conservation of momentum transverse to the brane.  This is taken into account in the above calculation by the inclusion of the image term in the scalar propagator.  It is the cross term between the ordinary scalar propagator and the image term in the second line of Eq.~(\ref{eq:int}) which leads to the brane localized logarithm in Eq.~(\ref{eq:result}).

Using Eq.~(\ref{eq:result}), the effective action becomes
\begin{eqnarray}
S_{eff} &=& S_{cl} - {m^4\over 256 \pi^2}\ln\left({\Lambda\over \mu}\right)\int d^4 x -{1\over 2!} {g_4 m^2\over 128\pi^2}\ln\left({\Lambda\over \mu}\right)\int d^4 x\phi(x^\mu,0)^2 \\
\nonumber
& & {} -{1\over 4!}{3 g_4^2\over 128\pi^2}\ln\left({\Lambda\over \mu}\right)\int d^4 x \phi(x^\mu,0)^4 + \cdots.
\end{eqnarray}
The logarithmic divergences in this expression can be absorbed into counterterms appearing in $S_{cl}.$  Using our prescription in which only powers of $\Lambda$ and $\ln(\Lambda/\mu)$ are subtracted, the RG equations for the brane couplings become
\begin{eqnarray}
\mu{d\lambda_0\over d\mu} &=& {m^4\over 256\pi^2},\\
\mu{d\lambda_2\over d\mu} &=& {\lambda_2^2\over \pi} + {m^2 g_4\over 128\pi^2},\\
\mu{d\lambda_4\over d\mu} &=& {4\lambda_4\lambda_2\over \pi} + {3 g_4^2\over 128\pi^2}.
\end{eqnarray}

There are also corrections from one-loop diagrams with insertions of both brane and bulk couplings.  First, consider the renormalization of the tension.  At linear order in $\lambda_2$ this is given by\
\begin{eqnarray}
G^{(0)} &=&
\begin{picture}(100,15)
\thicklines
\put(10,3){\circle*{4}}
\put(20,0){\makebox{$+$}}
\put(50,3){\circle{20}}
\put(40,3){\circle*{3}}
\put(70,0){\makebox{$+\cdots$}}
\end{picture}
\\
\nonumber 
&=& -\lambda_0(\Lambda)  - {1\over 2}\lambda_2(\Lambda)\int {d^4 k\over (2\pi)^4} D_k(0,0) +\cdots,
\end{eqnarray}
where we have included the effects of the bulk mass in the propagator
\begin{equation}
D_k(0,0)={1\over 2\pi} \ln \left({\Lambda^2\over k_4^2 + m^2}\right).
\end{equation}
To extract the RG equation for the tension, we need the coefficient of $\ln\Lambda$ in the tension counterterm.  We shall use 
\begin{equation}
\label{eq:in}
\int {d^4 k\over (2\pi)^4} D_k(0,0) = {m^4\over 64\pi^3}\ln \left({\Lambda^2\over \mu^2}\right) +\cdots,
\end{equation}
where finite terms, as well as terms proportional to powers of the cutoff or more powers of $\ln\Lambda$ have been suppressed.  Hence
\begin{equation}
G^{(0)} = -Z_0 \lambda_0(\mu) - {m^4\lambda_2\over 128\pi^3}\ln\left({\Lambda^2\over\mu^2}\right)+\cdots.
\end{equation}
Therefore, the one-loop beta function for the tension at ${\cal O}(\lambda_2)$ becomes
\begin{equation}
\mu {d\lambda_0\over d\mu} = {m^4\over 256\pi^2} - {m^4\lambda_2\over 64\pi^3}.
\end{equation}

We can also include brane couplings in the one-loop renormalization of $\lambda_2$ itself.  A similar calculation to the one above gives
\begin{equation}
\mu{d\lambda_2\over d\mu} = {\lambda_2^2\over\pi} - {m^4\lambda_4\over 64\pi^3}+ {m^2 g_4\over 128\pi^2},
\end{equation}
where for the one-loop part, we only included terms linear order in the couplings $\lambda_{2n}$ and $g_4$.  The pattern is similar for the other $\lambda_{2n}$ couplings.  

Finally, to complete the discussion of the RG flows in this theory, one should also calculate the beta functions for the bulk couplings.  Clearly, the brane couplings cannot generate bulk divergences, so these do not contribute. Then the calculation is a standard field theory exercise, which we will not repeat here.

\section{Conclusion}

In this paper, we have analyzed the types of ultraviolet divergences that appear in field theories with branes.  Non-dynamical branes induce a spacetime geometry that breaks translation invariance, leading to localized divergences at the quantum level.  Short distance divergences appear also in the limit of vanishing brane thickness.  Such divergences signify a failure of the theory to describe finite thickness effects, and manifest themselves already at the classical level.  By looking at a toy model with a 3-brane in six dimensions, we showed how to regulate and renormalize these classical singularities into the parameters of the theory, and derived RG equations (which are only generated in backgrounds with codimension-two branes) for the brane localized couplings.  We also computed corrections to these flows from one-loop bulk-to-brane effects.  Although we have worked with two non-compact extra dimensions, the divergences we encountered are only sensitive to short distance effects, and hence the RG equations we have derived remain valid if the space is compactified.

The brane-localized divergences considered here may have implications in the context of brane-world scenarios.  For example, models with two compact extra dimensions may address the hierarchy problem if the compact space is large~\cite{ADD}.  A mechanism for naturally generating a volume exponentially larger than the fundamental scale of the theory in such codimension-two models has been proposed by~\cite{arkani}.  This mechanism relies on large logarithms of the ratio of the size of the space over brane thickness induced by the bulk profile of a massless scalar that couples to 3-branes.  Bulk scalar induced classical logarithms also arise in the scenario of~\cite{arkani1} for obtaining gauge coupling unification with two large extra dimensions.  For a scalar field that is massless and non-interacting in the bulk, our classical RG flows could also play a role in this scenario, since they too generate such logarithms in the infrared.  

It may be interesting to see what happens when the brane is taken to be dynamical.  Also, it may be worthwhile to examine how a more fundamental description of the brane which includes finite thickness effects resolves the singularities and leads, at long wavelengths, to the classical running couplings described here.

\section{Acknowledgements}

This work was supported in part by the Department of Energy under grant number DE-FG03-92-ER 40701.

\appendix

\section{The Scalar Propagator on $R^6$ and $R^4\times R^2/Z_2$}

In this appendix we show the equivalence of the scalar propagator on cones of deficit angle $0$ ($\alpha=1$) and $\pi$ ($\alpha=1/2$) from Eq.~(\ref{eq:prop}) with the scalar propagator in flat six-dimensional space, and on the orbifold (given in Eq.~(\ref{eq:image}) as a sum over images), respectively.  We will need the Bessel function identities
\begin{eqnarray}
J_0(z) &=& \int_0^{2\pi} {d\theta\over 2\pi} e^{i z\cos\theta},\\
J_0(q R) &=& \sum_{n=-\infty}^\infty e^{i n\theta} J_n (q r) J_n (q r'),
\end{eqnarray}
with $R=\sqrt{r^2 + r'^2 - 2 r r'\cos\theta}.$  For $\alpha=1$, Eq.~(\ref{eq:prop}) gives
\begin{eqnarray}
D(x,x') &=& \int {d^4 k\over (2\pi)^4} \int_0^{\infty} {q dq\over 2\pi}{e^{i k_\mu (x^\mu-x'^\mu)} \over k_4^2 + q^2}\sum_n e^{in(\theta-\theta')} J_n(q r) J_n(q r')\\
\nonumber 
&=& \int {d^4 k\over (2\pi)^4} \int_0^{\infty} {q dq \over 2\pi}\int_0^{2\pi}{d\theta\over 2\pi}{e^{i k_\mu (x^\mu-x'^\mu)} e^{iq|{\vec x}-{\vec x'}|\cos\theta}\over k_4^2 + q^2},
\end{eqnarray}
or
\begin{equation}
D(x,x') = \int {d^4 k\over (2\pi)^4} \int {d^2 {\vec q}\over (2\pi)^2} {e^{i k_\mu (x^\mu-x'^\mu)+i{\vec q}\cdot ({\vec x}-{\vec x'})}\over k_4^2 + q^2},
\end{equation}
which is precisely the six-dimensional scalar propagator.  For $\alpha=1/2,$ Eq.~(\ref{eq:prop}) gives
\begin{equation}
\label{eq:a1/2}
D(x,x')=\sum_{n even}\int {d^4 k\over (2\pi)^4}\int_0^\infty {q dq\over \pi} {e^{ik_\mu (x^\mu-x'^\mu)} e^{in(\theta-\theta')}\over k_4^2 + q^2} J_n(q r) J_n(q r').
\end{equation}
We now write
\begin{eqnarray}
\sum_{n even} e^{in(\theta-\theta')}J_n(qr) J_n(qr') &=& {1\over 2}\left(\sum_{n even} + \sum_{n odd}\right) e^{in(\theta-\theta')}J_n(qr) J_n(qr')\\
\nonumber 
& & {}+{1\over 2}\left(\sum_{n even} - \sum_{n odd}\right) e^{in(\theta-\theta')}J_n(qr) J_n(qr')\\
\nonumber 
&=& {1\over 2}\sum_n e^{in(\theta-\theta')}J_n(qr) J_n(qr') + {1\over 2}\sum_{n} e^{in(\theta-\theta'+\pi)}J_n(qr) J_n(qr')\\
\nonumber
&=& {1\over 2} J_0(q|{\vec x}-{\vec x'}|) + {1\over 2} J_0(q|{\vec x}+{\vec x'}|).
\end{eqnarray}
Then Eq.~(\ref{eq:a1/2}) becomes 
\begin{equation}
D(x,x') = \int {d^4 k\over (2\pi)^4}\int {d^2{\vec q}\over (2 \pi)^2 } {e^{ik_\mu (x^\mu-x'^\mu)}\over k_4^2 + q^2} \left(e^{i {\vec q}\cdot ({\vec x}-{\vec x}')} + e^{i {\vec q}\cdot({\vec x} + {\vec x}')}\right),
\end{equation}
which reproduces Eq.~(\ref{eq:image}).

\section{Brane mass renormalization from classical field equations}

In the text, we obtained the RG equation for the brane mass $\lambda_2$ by renormalizing divergences in the diagrammatic expansion for the two-point function.  We now obtain the same beta function by solving the classical field equations for a free, massless bulk scalar with a brane localized mass terms.  This classical problem is also singular and must therefore be regularized.  The regulator dependence in the classical solution can be absorbed into a renormalized brane mass, which leads to the same result as Eq.~(\ref{eq:rg2}).

The full two-point function in the mixed representation is given by
\begin{equation}
G^{(2)}_k({\vec x},{\vec x}') = \sum_n\int_0^\infty {q dq\over k_4^2 + q^2}\phi_{n,q}^*({\vec x}')\phi_{n,q}({\vec x})
\end{equation}
where $\phi_{n,q}$ are a set of orthonormal functions which satisfy
\begin{equation}
\label{eq:modes}
\left(-\nabla_{\vec{x}}^2 + \lambda_2 \delta^2({\vec{x}})\right)\phi_{n,q}({\vec x}) = q^2 \phi_{n,q}({\vec x}).
\end{equation}
In polar coordinates, away from $r=0$ the solutions are, 
\begin{equation}
\phi_{n,q}({\vec x}) = {N_n(q)\over \sqrt{2\pi\alpha}} e^{in\theta/\alpha} R_{n,q} (r),
\end{equation}
where $N_n(q)$ is determined by normalization and $R_{n,q}(r)$ are linear combinations of Bessel functions
\begin{equation}
R_{n,q}(r) = J_{|n/\alpha|}(qr) + c_n(q)  Y_{|n/\alpha|}(qr),
\end{equation}
We obtain $c_n$ by applying the boundary conditions at $r=0$, which follow from integrating Eq.~(\ref{eq:modes}) over the interior of the surface $r=\epsilon$.  For the $n\neq 0$ modes, this integral vanishes due to rotational invariance.  Thus $\lambda_2$ has no effect on these modes and we conclude that $c_n=0$ for $n\neq 0$.  For the $n=0$ mode, we get from Eq.~(\ref{eq:modes})
\begin{equation}
\label{eq:bound}
-\epsilon\left.{dR_{0,q}\over dr}\right|_{r=\epsilon} + {\lambda_2\over 2 \pi\alpha} \int_0^\epsilon dr \delta(r) R_{0,q}(r) - q^2 \int_0^\epsilon r dr R_{0,q}(r)=0,
\end{equation}
where we used $\delta^2({\vec x}) = \delta (r)/2\pi\alpha r$.  The second term in this equation is singular as $\epsilon\rightarrow 0$, due to the singularity of $Y_0$ at the origin.  We will handle this singularity by regulating the delta function:
\begin{equation}
\delta(r) = {1\over \delta}\left[1-\theta(r-\delta)\right], 
\end{equation}
with $\delta\rightarrow 0.$  Using the asymptotic form for the Bessel functions near $r=0$,
\begin{equation}
R_{0,q}(r) \simeq 1 +{2c_0(q)\over\pi}\left[\gamma + \ln\left({q r\over 2}\right)\right],
\end{equation}
we find that the third term in Eq.~(\ref{eq:bound}) vanishes as $\epsilon\rightarrow 0$.  On the other hand,
\begin{equation}
\epsilon \left.{d R_0\over dr}\right|_{r=\epsilon} ={2 c_0(q)\over \pi}
\end{equation}
and using the regulated form for the delta function (with $\epsilon>\delta$),
\begin{eqnarray}
\int_0^\epsilon dr \delta (r) R_0(r) &=& {1\over \delta}\int_0^\delta dr R_0(r)\\
\nonumber 
 &=& 1 + {2 c_0(q)\over \pi} \left[\gamma -1 + \ln\left({q\delta\over 2}\right)\right].
\end{eqnarray}
Then Eq.~(\ref{eq:bound}) gives
\begin{equation}
{2c_0(q)\over\pi} = {\lambda_2/(2\pi\alpha)\over 1 -(\lambda_2/2\pi\alpha)\ln (q/\Lambda)},
\end{equation}
where $1/\Lambda = e^{\gamma-1}\delta /2$.  As in the text, the dependence of $c_0(q)$ on the regulator can be removed by interpreting $\lambda_2$ as a bare coupling $\lambda_2(\Lambda)$ and introducing a renormalized parameter $\lambda_2(\mu)=\lambda_2(\Lambda)/Z_2$.  In terms of $\lambda_2(\mu)$,
\begin{equation}
{2c_0(q)\over\pi} = {\lambda_2(\mu)/(2\pi\alpha)\over 1 -(\lambda_2(\mu)/2\pi\alpha)\ln (q/\mu)},
\end{equation}
provided that $1/Z_2 = 1-(\lambda_2(\mu)/2\pi\alpha)\ln (\Lambda/\mu)$.  This gives
\begin{equation}
\mu {d\lambda_2\over d\mu} ={\lambda_2^2\over 2\pi\alpha},
\end{equation}
in agreement with the diagrammatic approach.

\end{document}